\documentclass{article}
\usepackage{spconf}
\usepackage{subcaption}
\usepackage{amsmath,amssymb}
\usepackage{hyperref}
\usepackage{url}
\usepackage{booktabs}       
\usepackage{amsfonts}       
\usepackage{nicefrac}       
\usepackage{microtype}      
\usepackage{xcolor}         
\usepackage[pdftex]{graphicx}
\usepackage[numbers, sort]{natbib}
\usepackage{tikz}
\usepackage{pgfplots, pgfplotstable}
\usepackage{enumitem}
\pgfplotsset{compat=1.18}
\usetikzlibrary{calc, shapes}
\usepackage{arydshln}
\usepackage{setspace}
\usepackage{natbib}

\usepackage{xspace}
\makeatletter
\DeclareRobustCommand\onedot{\futurelet\@let@token\@onedot}
\def\@onedot{\ifx\@let@token.\else.\null\fi\xspace}
 
\def\ie{\emph{i.e}\onedot}

\def\etal{\emph{et al}\onedot}
\makeatother

\title{Learning Semantic Information from Raw Audio Signal \\ Using Both Contextual and Phonetic Representations}
\name{Jaeyeon Kim$^{1,2 \star}$, Injune Hwang$^{3}$, Kyogu Lee$^{3}$\thanks{$^{\star}$This work is the continuation of the work done during the internship in Music \& Audio Research Group, Seoul National University.}} 
\address{
  $^{1}$MAUM AI Inc., Republic of Korea, \\ $ ^{2}$Seoul National University, Republic of Korea 
  \\ $^{3}$Music \& Audio Research Group (MARG), Seoul National University, Republic of Korea 
}
\begin{document}
\maketitle
\begin{abstract}
We propose a framework to learn semantics from raw audio signals using two types of representations, encoding contextual and phonetic information respectively. Specifically, we introduce a speech-to-unit processing pipeline that captures two types of representations with different time resolutions. For the language model, we adopt a dual-channel architecture to incorporate both types of representation. We also present new training objectives, masked context reconstruction and masked context prediction, that push models to learn semantics effectively. Experiments on the sSIMI metric of Zero Resource Speech Benchmark 2021 and Fluent Speech Command dataset show our framework learns semantics better than models trained with only one type of representation.
\end{abstract}

\section{Introduction}
\vspace{-0.3\baselineskip}
Spoken language models (SLMs) are language models (LMs) trained with raw audio signals without any text
\cite{nguyen2020zero}. 
Common approaches to spoken language modeling \cite{nguyen2020zero, lakhotia-etal-2021-generative} consist of the following components: a speech representation model trained without text, a quantization module, and an LM. The first two components are often jointly referred to as speech-to-unit component \cite{lakhotia-etal-2021-generative, textless-lib}. The speech-to-unit component transforms an acoustic signal into a sequence of discrete units, and the LM is trained on this pseudo-text. 

Nguyen \etal \cite{nguyen2020zero} introduced spoken language modeling and  Zero Resource Speech Benchmark (ZRSB) 2021, which evaluates the phonetic, lexical, syntactic, and semantic aspects of SLMs. 
Their work also included baseline models: pretrained CPC \cite{oord2018representation} for speech representation, k-means for quantization, BERT \cite{devlin-etal-2019-bert} or LSTM for LM. 
Subsequent studies attempted to improve baseline models by enhancing speech representations \cite{niekerk21_interspeech, chorowski21_interspeech, maekaku21_interspeech, cpc_deep_cluster, hubert_zerospeech, vg_baseline, fast-vgs-zero}. Lakhotia \etal proposed Generative Spoken Language Modeling (GSLM), generating speech with a generative LM trained with quantized CPC, wav2vec2.0 \cite{wav2vec2}, and HuBERT \cite{hubert}.
Speech generation using SLMs has further advanced to model prosodic information \cite{prosody_gslm} or generate spoken dialogues \cite{dialogue_gslm}.

Previous works showed the feasibility of learning a language from raw audio signals. Among different linguistic aspects, learning the semantic aspect still remains the most challenging task \cite{dunbar21_interspeech, hubert_zerospeech}. 
We hypothesize that the speech representations used in previous works mainly encode phonetic information, and giving more contextual information together will improve semantic learning. 

In this paper, we propose a framework to train SLMs using both contextual and phonetic representations. 
We also introduce masked context reconstruction (MCR) and masked context prediction (MCP) tasks to learn semantics from both representations.
Our framework allows models to learn semantics more effectively compared to models that utilize only one type of representation

\begin{figure*}[t!]
    \centering
    \includegraphics[width=0.95\textwidth]{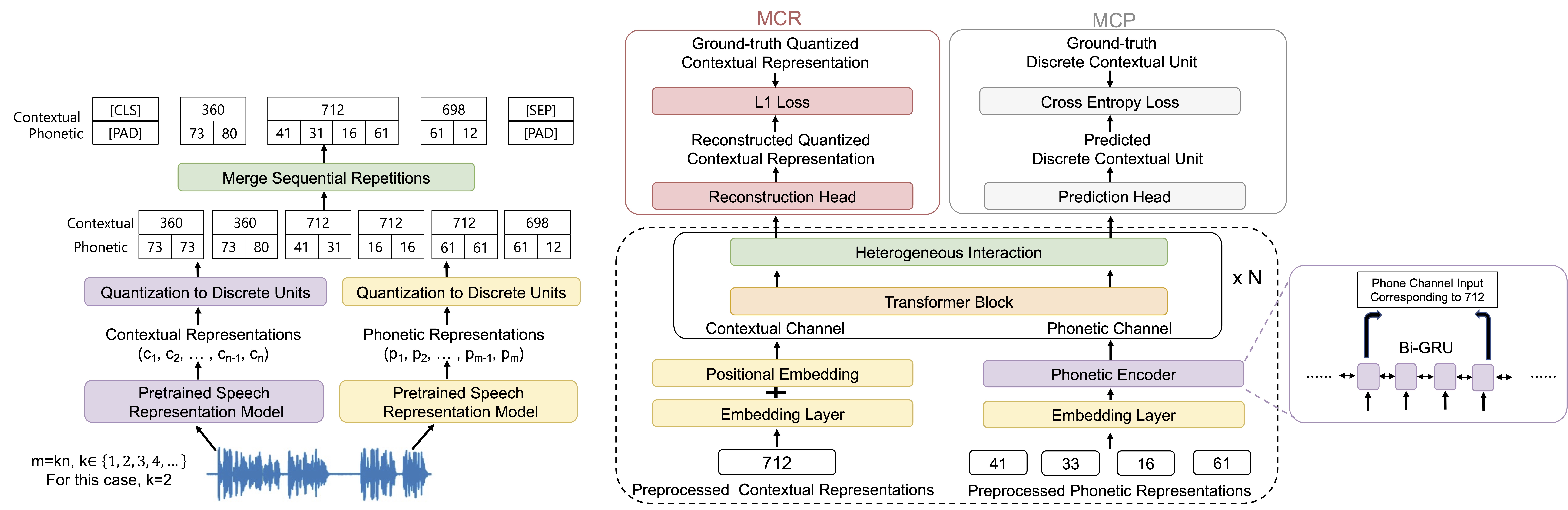}
        \vspace{-0.7\baselineskip}
    \caption{Example of speech-to-unit processing and the overall architecture of our framework. Integer values denote the discrete contextual and phonetic units}
    \label{fig:processing_architecture}
    \vspace{-1\baselineskip}
\end{figure*}

\vspace{-0.3\baselineskip}
\section{Proposed Framework}

\vspace{-0.3\baselineskip}
\subsection{Speech-to-Unit Processing}\label{preprocessing}
The audio signal first passes through two pretrained speech representation models, chosen based on the constraint that the time interval between representations of one model is a multiple of the time interval between other's.
We refer to the representations with lower resolution and thus encode relatively contextual information as contextual representations, while the others as phonetic representations. Additionally, we attempt to select representations encoding more contextual information than phonetic details, such as word meaning or word identity, as contextual representations. 

Both contextual and phonetic representations are quantized to discrete units, \ie the codebook indices, and then aligned based on time intervals between the representations. 
As shown in Figure \ref{fig:processing_architecture}, if the time interval between contextual representations is twice longer, two phonetic units are matched to one contextual unit. The aligned units are then merged to remove sequential repetitions, as in GSLM \cite{lakhotia-etal-2021-generative}. 
Repeated contextual units are always merged, while repeated phonetic units are not merged if they are aligned with different contextual units. We add \textit{[CLS]} and \textit{[SEP]} to the contextual units, while adding padding as corresponding phonetic units.

\vspace{-0.5\baselineskip}
\subsection{Language Model Architecture}
Many existing models in the audio domain use BERT with heterogeneous inputs by joining heterogeneous sequences along the sequential dimension \cite{png_bert, phoneme_bert, st-bert, speechbert},
but this approach requires much more computation and memory due to the longer input sequence. In addition, it can be difficult to discern which contextual and phonetic representations correspond to the same time segment of speech in a joint sequence. Therefore, we adopted a dual-channel architecture of contextual and phonetic channels. The architecture is based on CharBERT \cite{ma-etal-2020-charbert}, which was originally designed to address the challenges of subword tokenization.
The overall architecture is depicted in Figure \ref{fig:processing_architecture}.

\noindent
\textbf{Embedding Layer.} Contextual and phonetic units are first embedded by separate embedding layers. 
Then embedded contextual units are given as the input of the contextual channel. 

\noindent
\textbf{Phonetic Encoder.} The phonetic encoder is a bidirectional GRU layer with a hidden dimension that is half of the transformer block's. It operates on the entire sequence of embedded phonetic units. The outputs of the first and last units that align with a merged contextual unit are concatenated to form the phonetic channel input. 

\noindent
\textbf{Dual-Channel Transformer Block.}
For each transformer block, contextual channel representations first pass through the block, and self-attention is calculated solely among these representations. Then, phonetic channel representations pass through the identical transformer block in the same way. 

\noindent
\textbf{Heterogeneous Interaction.} 
The heterogeneous interaction module fuses and splits the hidden states of contextual and phonetic channels to enrich each other after every transformer layer. The hidden states of contextual and phonetic channels first pass through separate linear layers, then concatenated and fused by 1D convolution. The fused representations are split into representations for each channel again by passing through another separate linear layer. The residual connection is applied to separated representations to maintain the information from each channel.

\vspace{-0.5\baselineskip}
\subsection{Training Task}
We train our framework on two self-supervised tasks utilizing masking. We mask contextual units while leaving phonetic units intact. When two tasks are employed together, the total loss for training is the summation of the losses of each task. 

\noindent
\textbf{Masked Context Reconstruction (MCR).} We reconstruct quantized contextual representations, \ie codebook entries, corresponding to masked contextual units using L1 loss. The transformer outputs of the contextual channel for the masked input are used for the reconstruction

\noindent
\textbf{Masked Context Prediction (MCP).}  We predict masked contextual units with cross entropy loss. When training with MCP only, the transformer outputs of the contextual and phonetic channels corresponding to the masked input are concatenated and used for prediction, while only the outputs of the phonetic channel are used when trained with both MCR and MCP.

Prior works \cite{prosody_gslm, discrete_unit} have shown that the discretization of speech representations disentangles the linguistic content from other acoustic information such as prosody and speaker information. In line with this observation, MCP focuses on acquiring contextualized information related to linguistic contents encoded in contextual units. This is similar to masked language modeling (MLM) \cite{devlin-etal-2019-bert}, while MCP additionally aims to capture the relationship between contextual and phonetic units via utilizing unmasked phonetic units for prediction. On the other hand, MCR aims to learn acoustic information that may be lost during discretization but could contribute to semantic learning. To achieve this, MCR reconstructs the representations before the discretization process. 

\vspace{-0.3\baselineskip}
\section{Experiment}
\vspace{-0.3\baselineskip}
\subsection{Implementation of the Framework}
\noindent
\textbf{Contextual and Phonetic Representations.}
We try to apply our framework to diverse speech representations.
From visually grounded speech representation models, ResDAVEnet-VQ (RDVQ) \cite{ResDAVEnet-VQ}, known to learn hierarchical linguistic units, is chosen. 
From self-supervised speech representation models, CPC \cite{oord2018representation}, wav2vec2.0 \cite{wav2vec2}, and HuBERT \cite{hubert} are chosen following GSLM \cite{lakhotia-etal-2021-generative}.
For RDVQ, we only use the audio encoder of the pretrained model. 

Let VQ2 and VQ3 denote the representations extracted from the 2nd and 3rd vector quantization (VQ) layer of RDVQ. VQ3 and wav2vec2.0 representations from intermediate layers are chosen as contextual representations since they are related to the word meaning \cite{ResDAVEnet-VQ, layerwise_analysis}. VQ2, CPC, and HuBERT are chosen as phonetic representations based on their strong phonetic performance \cite{ResDAVEnet-VQ, lakhotia-etal-2021-generative}. The details of the chosen representations are given in Table \ref{tab:representaion_details}. 

\noindent
\textbf{Quantization.} VQ3 and VQ2 are naturally quantized by VQ layers of RDVQ, with a codebook size of 1024. Others are quantized using k-means. The codebook size for CPC and HuBERT are 100 based on the observations from GSLM \cite{lakhotia-etal-2021-generative}.
However, as the number of contexts far exceeds the number of phones, we attempt to use a much larger codebook size of 1024 for wav2vec2.0 following VQ3

\setcounter{table}{0}
\begin{table}
    \caption{The details of phonetic and contextual representations. Layer denotes from which layer of the model the representations were extracted. VQ2 and VQ3 denote the second and third vector quantization layer of the model respectively.}
    \vspace{-0.5\baselineskip}
    \centering
    \resizebox{0.9\linewidth}{!}{
    \begin{tabular}{ccccc}
    \hline
    Representation & Model & Layer & \multicolumn{1}{l}{Interval} & Codebook Size \\ \hline
    \multicolumn{5}{c}{\textbf{Contextual Representation}}  \\              
    VQ3 & RDVQ & VQ3 & 40ms & 1024 \\
    wav2vec2.0 & wav2vec2.0 Base & 7 & 20ms & 1024 \\ \hline
    \multicolumn{5}{c}{\textbf{Phonetic Representation}}  \\            
    VQ2 & RDVQ & VQ2 & 20ms & 1024 \\
    CPC & CPC  & last & 10ms & 100 \\
    HuBERT & HuBERT Base & 6 & 20ms & 100 \\
    \hline
    \end{tabular}
    }
    \label{tab:representaion_details}
    \vspace{-0.5\baselineskip}
\end{table}

\noindent
\textbf{Model Configuration.}
The implementation of Transformer blocks is based on BERT-small \cite{nguyen2020zero}, which has 8 hidden layers with 512 hidden dimensions and 8 attention heads, having 27M parameters in total. We remove one hidden layer for our framework to match the number of parameters, as the phonetic encoder, the heterogeneous interaction module, and the heads for MCP and MCR tasks require additional parameters.

\setcounter{table}{1}
\begin{table*}
    \caption{Overall sSIMI scores of the models. libri. and synth. denote the sSIMI score for the LibriSpeech subset and the synthesized subset, respectively. The Model column denotes the representation used, or the reference of the model.}
    \label{experiment_result}
            \vspace{-0.5\baselineskip}
    \centering
    \resizebox{0.85\textwidth}{!}{
    \begin{tabular}{ccc:cc|cccc}
        \hline\
        Model & libri. ($\uparrow$) & synth. ($\uparrow$) & libri. ($\uparrow$) & synth. ($\uparrow$). &
        Model & libri. ($\uparrow$) & synth. ($\uparrow$) &  GPU hour \\
        \hline 
        & \multicolumn{2}{c:}{\textbf{\textit{(a) Pretrained}}}  & \multicolumn{2}{c|}{\textbf{\textit{(b) BERT-small}}} & wav2vec2.0 + CPC + MCP & \textbf{13.2} & \textbf{13.14} & 60 \\ 
        \cdashline{6-9}
        & \multicolumn{2}{c:}{\textbf{\textit{Representations}}} &  \multicolumn{2}{c|}{} & \multicolumn{4}{c}{\textbf{\textit{Ablation}}} \\
        VQ3 & 5.17 & -1.50 & 9.82 & 3.69 & w/o heterogeneous interaction & 11.23 & 10.69 & 60\\
        wav2vec2.0 & 3.35 & 2.46 & 9.73 & 10.91 &  w/o phonetic encoder & 11.56 & 12.65 & 60 \\
        VQ2 & 5.56 & 2.61 &  8.16 & 6.38 & BERT-small with joint sequence & 9.93 & 10.80 & 60\\
        \cdashline{6-9}
        CPC & 12.04 & 0.40 & 7.13 & 5.33 & \multicolumn{4}{c}{\textbf{\textit{Comparison to Other Spoken Language Models}}}\\
        HuBERT & 5.17 & 4.11 & 11.73 & 8.42 &CPC-big + BERT-small \cite{nguyen2020zero}  & 5.56 & 3.88 & 60 \\
        Average & 6.26 & 1.62 & 9.32 & 6.95 & CPC-big + LSTM \cite{nguyen2020zero} & 7.56 & 4.42 & 60  \\ 
        \cdashline{0-4} 
        & \multicolumn{4}{c|}{\textbf{\textit{Our Framework}}} & van Niekerk \etal \cite{dunbar21_interspeech, niekerk21_interspeech}& 7.69 & 4.29 & 60 \\        
        & \multicolumn{4}{c|}{\textbf{\textit{trained with}}} & Chorowski \etal \cite{dunbar21_interspeech, chorowski21_interspeech}& 10.2 & 5.9 & 60 \\
        & \multicolumn{2}{c:}{\textbf{\textit{(c) MCP}}} & \multicolumn{2}{c|}{\textbf{\textit{(d) MCR and MCP}}} & Maekaku \etal \cite{dunbar21_interspeech, maekaku21_interspeech}& 8.89 & -2.10 & 60\\
        VQ3 + VQ2 & 12.81 & 10.24 & 13.92 & 8.74 & Liu \etal \cite{dunbar21_interspeech} & 3.16 & 1.79 & 60 \\
        VQ3 + CPC & 9.20 & 9.53 & 9.84 & 10.08 & Krishna \etal \cite{cpc_deep_cluster} & 10.25 & 3.96 & 60 \\
        VQ3 + HuBERT & 13.83 & 9.66 & 14.70 & 9.33 & Maekaku \etal \cite{hubert_zerospeech} & 12.53 & 2.86 & 60 \\
        wav2vec2.0 + VQ2 & 11.25 & 13.01 & \textbf{15.61} & 7.89 &  Alishahi \etal \cite{vg_baseline} & 12.61 & 9.65 & 72 \\
        wav2vec2.0 + CPC & 13.2 & 13.14 & 11.00 & \textbf{13.60} & \multicolumn{4}{c}{Topline Models \cite{nguyen2020zero}} \\
        wav2vec2.0 + HuBERT & 8.69 & 12.20 & 12.74 & 9.05 & Forced align BERT & 4.54 & 7.92 & 1536 \\
        Average & 11.50 & \textbf{11.30} & \textbf{12.97} &  9.78 & Phone BERT & 16.11 & 9.86 & 1536 \\
        \hline
    \end{tabular}}
    \vspace{-0.8\baselineskip}
\end{table*}
\vspace{-0.5\baselineskip}
\subsection{Experiment on sSIMI}\label{sSIMI_experiment}
\textbf{sSIMI metric. }We use the sSIMI similarity metric of the ZRSB 2021 \cite{nguyen2020zero} to evaluate the semantic learning of the models.
Let $x$ and $y$ denote the input representations corresponding to audio files of the pair of words.
The semantic distance $d$ between two audio files given by the model is computed as 
\begin{align}
    d(x, y) = \mathrm{dist}(&f_\mathrm{pool}(h^{(i)}(x)), f_\mathrm{pool}(h^{(i)}(y))) ,
    \label{eq:sSIMI}
\end{align}
where $f_\mathrm{pool}$ is the pooling function, $h^{(i)}$ is the output of $i$th hidden layer of the model, and $\mathrm{dist}$ is the distance function such as 
euclidean distance. 
The sSIMI metric is defined as Spearman's rank correlation between the semantic distance given by the model and the human for the pairs of words.  
A higher score means that the model's judgment of the semantic similarity between word pairs aligns more closely with human judgment, indicating better learning of lexical semantics.

We evaluate the sSIMI score on the development set, which includes
audio files corresponding to English word pairs from mturk-771 \cite{mturk_771}. The development set comprises two subsets: the LibriSpeech subset created by extracting audios from LibriSpeech \cite{librispeech}, and the synthesized subset created by synthesizing audios using Google API. 
We evaluate overall performance based on the summation of the scores for both subsets. 

\noindent\textbf{Setup.} We evaluate sSIMI scores of representations in Table \ref{tab:representaion_details} for following setups.
\vspace{-1.5mm}
\begin{itemize}[leftmargin=0.45cm]
    \setlength\itemsep{-1mm}
    \item (a): Pretrained representations themselves.
    \item (b): BERT-small trained with one type of representation. Models are trained with only MLM, and the input representations are quantized and merged as in Sec. \ref{preprocessing}. 
    \item Our framework trained with (c): only MCP and \\ (d): both MCR and MCP. 
\end{itemize}
\vspace{-1.5mm}

All models are trained on the audio files from LibriSpeech 960hours \cite{librispeech}. Every model was trained with an NVIDIA A100 GPU for 60 hours with a batch size of 16. We used AdamW optimizer and Transformer learning rate scheduler with linear warm-up over 10000 steps. We used the masking probability of 15\% for masking-based tasks.

\noindent\textbf{Result. } The overall result is demonstrated in Table \ref{experiment_result}.

\textbf{\textit{Language Modeling.}} sSIMI scores of (b) are generally higher than (a), indicating that language modeling benefits semantic learning.

\textbf{\textit{Effectiveness of Our Framework.}} sSIMI scores of (c), (d) are generally higher than the corresponding models of (b), despite being trained with similar budgets. 
This demonstrates the effectiveness of our framework in semantic learning, and this holds for different combinations of representations. Furthermore, considering that MCP and MLM are similar prediction tasks except that MCP incorporates phonetic information for prediction, high scores of (c) suggest that utilizing both contextual and phonetic information benefits semantic learning.

\textbf{\textit{MCR and MCP.}} sSIMI scores of (d) are generally higher on the LibriSpeech subset than (c). This indicates that additionally utilizing acoustic information encoded in contextual representations through the MCR can help learn more specific semantic information of the training dataset. However, (d) scores lower on the synthesized subset than (c), suggesting that learning more specific semantics for a particular dataset can lead to a degradation of performance on other datasets. 
Note that the overall scores of (c) and (d) are nearly identical.

\textbf{\textit{Ablation.}} We conducted ablation experiments on the architecture with the model using wav2vec2.0 and CPC with only MCP. 
We compared the model to BERT-small trained on a joint sequence of wav2vec2.0 and CPC.
Results in Table \ref{experiment_result} show that our framework based on dual-channel architecture is more effective for semantic learning than the joint sequence approach when both contextual and phonetic information is utilized. Furthermore, both the heterogenous interaction and the phonetic encoder contribute to semantic learning when adopting the dual-channel architecture. 

\textbf{\textit{Comparison to Other SLMs.}}
We compare our best model, which uses wav2vec2.0 and CPC with only MCP, to similar-budget SLMs trained with improved CPC \cite{niekerk21_interspeech, chorowski21_interspeech, maekaku21_interspeech, cpc_deep_cluster}, improved HuBERT \cite{hubert_zerospeech}, and visually grounded speech representations \cite{vg_baseline}.
Our model outperforms all other models while using na\"ive wav2vec2.0 and CPC. Note that improvements of representations can be incorporated into our framework by using improved representations as contextual or phonetic representations. 
We further compare our best model with topline models from ZRSB 2021 \cite{nguyen2020zero}. Our model significantly outperforms BERT (90M parameters) trained with force-aligned phonemes of LibriSpeech and performs comparably with BERT trained with gold phonetic transcriptions of LibriSpeech. Note that our models are trained with a much smaller budget, and only with raw audio signals. 

\vspace{-0.5\baselineskip}
\subsection{Experiment on Spoken Language Understanding}
\noindent
\textbf{Setup.} We evaluate our framework on the Fluent Speech Command (FSC) dataset \cite{fsc}, one of the most widely used SLU dataset. The task is to classify the intent of the given speech. 
We additionally train and evaluate splits of FSC from Arora \etal \cite{fsc_mase}, composed of an unseen-speaker split to test speech processing and an unseen-utterance split to test semantic processing.  
Models using wav2vec2.0 and CPC trained in Sec \ref{sSIMI_experiment} are additionally trained for the following setups: \textit{(e) directly finetuning on FSC for 2 hours, (f) pretraining with MLM, MCP, and MCR on FSC for 1 hour followed by 2 hours of finetuning}. For finetuning, we add a linear classifier over the \textit{[CLS]} token.

\setcounter{table}{2}
\begin{table}[]
\caption{Test Accuracy on Fluent Speech Command.}
\label{fsc_result}
\vspace{-0.5\baselineskip}
\label{metric}
\centering
\resizebox{0.85\linewidth}{!}{
\begin{tabular}{ccccc}
\hline
\multicolumn{2}{c}{} & & \multicolumn{2}{c}{Unseen} \\
\multicolumn{2}{c}{Model} & Original& Speaker& Utterance \\
\hline
\multicolumn{5}{c}{\textbf{\textit{(e) Direct finetuning}}} \\
\multicolumn{2}{c}{wav2vec2.0 + BERT-small} & 94.67\% & 90.49\% & 74.69\% \\
\multicolumn{2}{c}{CPC + BERT-small} & 65.49\% & 45.13\% & 34.40\% \\
\multicolumn{2}{c}{Ours + MCP} & 96.12\% & \textbf{91.56\%} & 80.31\%  \\
\multicolumn{2}{c}{Ours + MCP \& MCR} &\textbf{96.26\%}& 91.27\% & \textbf{81.84\%} \\
\cdashline{0-4}
\multicolumn{5}{c}{\textbf{\textit{(f) Addtional Pretraining}}} \\
\multicolumn{2}{c}{wav2vec2.0 + BERT-small} & 94.93\% & 90.97\% & 74.49\% \\
\multicolumn{2}{c}{CPC + BERT-small} & 67.91\% & 63.01\% & 50.89\% \\
\multicolumn{2}{c}{Ours + MCP} & 96.18\% & 91.53 \% & 81.99\%  \\
\multicolumn{2}{c}{Ours + MCP \& MCR} &\textbf{96.76\%}& \textbf{91.80}\% & \textbf{84.16\%}\\
\hline
\end{tabular}
}
\vspace{-\baselineskip}
\end{table}
\noindent\textbf{Results.} The overall results are presented in Table \ref{fsc_result}. Models based on our framework achieved higher accuracy than BERT-small trained with one type of representation. The performance gap is particularly significant for the unseen-utterance split, showing that the effectiveness of our framework in semantic learning also holds for the SLU task. Models of (f) generally outperform models of (e), suggesting that pretraining tasks of SLMs help to learn semantic information that is useful for SLU. Notably, for models using wav2vec2.0, pretraining with MCR shows the greatest improvements in accuracy. This indicates that utilizing acoustic information encoded in contextual representations through 
MCR helps to learn more specific semantics of the training dataset, as observed in Sec \ref{sSIMI_experiment}.

\vspace{-0.3\baselineskip}
\section{Conclusion}
\vspace{-0.5\baselineskip}
In this work, we proposed a framework that utilizes both contextual and phonetic representations to train more semantic-aware models with raw audio signals. Results on the sSIMI metric showed that our framework is effective in learning lexical semantics and is compatible with various representations. Experiments on FSC demonstrated that the advantage of our framework also holds for spoken language understanding. We expect future works to employ our framework in conjunction with improvements in speech representations 
to further advance semantic learning.

\vspace{-0.3\baselineskip}
\section{Acknowledgement}
\vspace{-0.5\baselineskip}
We would like to thank Junhyeok Lee from Supertone, Minho Kim from MIT, and Sang Hoon Woo for valuable discussions. 
This work was supported by Institute of Information \& communications Technology Planning \& Evaluation (IITP) grant funded by the Korea government(MSIT) (No.2022-0-00641).
\vfill\pagebreak
\begin{spacing}{0.95}
\bibliographystyle{IEEEbib}
\setlength{\bibsep}{2pt}
\bibliography{main}
\end{spacing}

\end{document}